\documentclass[12pt,preprint]{aastex}

\usepackage{natbib}
\usepackage[normalem]{ulem} 

\slugcomment{}

\shorttitle{Horizontal Flows in Active Regions}
\shortauthors{Jain et al.}

\begin{document}


\title{Horizontal Flows in Active Regions from Ring-diagram 
and \\Local Correlation Tracking Methods}

\author{Kiran Jain and S.C. Tripathy}
\affil{National Solar Observatory, 950 N Cherry Av., Tucson, AZ 85719, USA}
\email{kjain@nso.edu}

\author{B. Ravindra}
\affil{Indian Institute of Astrophysics, Block 2, Koramangala, Bangaluru, 560034, India}

\author{R. Komm and F. Hill}
\affil{National Solar Observatory, 950 N Cherry Av., Tucson, AZ 85719, USA}

\begin{abstract}

Continuous high-cadence and high-spatial resolution Dopplergrams allow us to study sub-surface 
dynamics that may be further extended to explore precursors of visible solar activity on the 
surface. Since the $\it p$-mode power is absorbed in the regions of high magnetic field, the inferences 
in these regions are often presumed to have large uncertainties. In this paper, using the Dopplergrams 
from space-borne Helioseismic Magnetic Imager  (HMI), we compare horizontal flows 
in a shear layer below the surface and the photospheric layer  
in and  around active regions. The photospheric 
flows are calculated using local correlation tracking (LCT) method while the ring-diagram 
(RD) technique of helioseismology is used to infer flows in the sub-photospheric shear layer. We find a strong positive 
correlation between flows from both methods near the surface. This implies that despite 
the absorption of acoustic power in the regions of strong magnetic field, the flows inferred
from the helioseismology are comparable to those from the surface measurements. However, the magnitudes 
are significantly different; the flows from the LCT method are 
smaller by a factor of 2 than the helioseismic measurements. Also, the median difference between
direction of corresponding vectors is $49\degr$.

\end{abstract}


\section{Introduction}
The dynamic nature of the Sun is manifested by eruptions observed in different layers of the solar
atmosphere. These eruptions generally originate from regions that harbor strong magnetic fields.
Propagating acoustic waves interact with these regions and modify their properties.  It is believed that the presence
of high magnetic field  affects the equilibrium profile of sound speed and density in the solar
interior that further modifies the characteristics of acoustic oscillations. As a result, while 
oscillation frequencies vary in phase with the solar activity cycle \citep[e.g.,][]{jain09}, the acoustic power is anti-correlated \citep{rajaguru01, jain08}. 
The availability of high-spatial resolution 
continuous Doppler images in the last two decades have opened a new dimension in the field of helioseismology
where properties of active regions below the solar surface are being explored in detail. However, these studies
are often subjected to the uncertainties in inferred properties, such as sub-surface structure and 
flows, due to the absorption of acoustic power.

The sub-surface flows in active regions are estimated using the techniques of local helioseismology. These
techniques are capable of probing the solar interior in three dimensions  \citep{antia07} and 
allow us to infer flows in different layers from the surface to several Mm in depth which
have become a crucial ingredient in computing solar dynamo models. While the ring--diagram 
method  \citep[RD;][]{RD} provides flows in shallow layers \citep[e.g.,][]{irene06,jain12, jain15, komm15a}, 
the time--distance 
method \citep[TD;][]{TD} is capable of providing flows in much deeper layers \citep[e.g.,][]{zhao13, kholikov14}. 
A detailed comparison between the horizontal flows obtained from these two methods was carried out 
by \citet{brad04} using data from two years of the  Michelson Doppler Images Dynamics Program. These authors 
found a good agreement between flows obtained from RD and TD methods. 

On the other hand,
efforts have also been made to validate flows near the surface from helioseismology by directly
comparing with surface measurements, mainly with feature tracking methods. \citet{svanda07} compared
the  flows obtained from TD and local correlation tracking \citep[LCT;][]{november88} methods and reported a significantly high
correlation, however the magnitudes from both methods were significantly different. They suggested that the
measurements from TD were correct and the magnitude of the LCT measurements must be corrected.
Measurements from the TD and
LCT methods were also compared with the realistic simulations of solar convection by \citet{dali07} where
authors found similar large-scale convective patterns. This study was primarily confined to using the $f$-mode ridge.
Later, \citet{zhao07} validated the TD method by computing acoustic travel times and inferring mean
flow fields using a ray approximation based inversion  for different depths of the simulated data of \citet{dali07}.
More studies were carried out recently using the Helioseismic and Magnetic Imager (HMI) observations to verify 
helioseismic methods 
\citep{liu13,svanda13}, however none was focused at validating flows from the RD method. 
In all these studies, strong positive correlations were obtained for both zonal and meridional components.
In addition,  \citet{liu13} have also identified the areas where the angle between the flows from TD and DAVE4VM was greater 
than $90\degr$, i.e., the flows point to the different directions. Most of these areas were either inside the sunspot penumbra or far away from the sunspot
where the DAVE4VM becomes insensitive due to the weak magnetic field strength.
In this paper,
we present results on the horizontal flow measurements from RD and LCT methods with several aims;
(i) validation of horizontal flows from inversion in the RD method, (ii) how different/similar are the flows from
helioseismic and surface measurements,  (iii) identify areas where both methods diverge or converge, 
and (iv) does the helioseismic measurements
provide reliable estimates of flows in the regions of high-magnetic field.

The paper is organized as follows: the selection of data and the methods to derive
 horizontal flows are described in Section 2. In Section
3, we discuss and compare results obtained from different methods and finally, the findings
of this study are summarized in Section 4. 

\section{Selection of Data and Estimation of Flows}
\begin{figure*}   
   \centerline{
\includegraphics[scale=1.2]{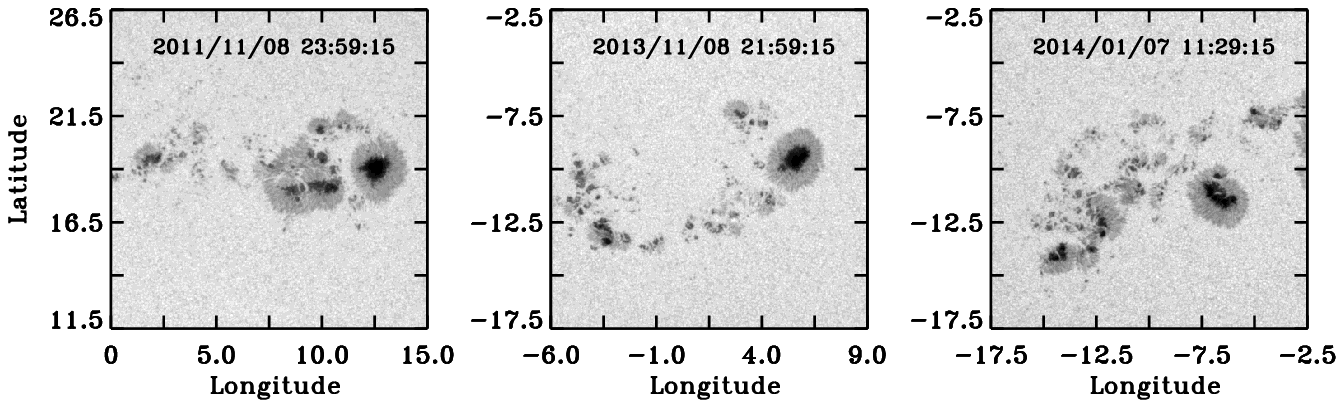}
}
            \caption{HMI Continuum images of three active regions used in this study ; (left) AR 11339,
           (middle) AR 11890, and (right) AR 11944.}
   \label{regions}
   \end{figure*}

We select three active regions from the current solar cycle; NOAA 11339, 11890 and 11944, and use 
45 s cadence continuous high-spatial resolution full-disk
Dopplergrams from the HMI \citep{hmi} onboard  {\it Solar Dynamics Observatory} \citep[SDO;][]{sdo} to calculate flows
in the regions of high magnetic field. Images of all three active regions in the HMI continuum  are shown in 
Figure~\ref{regions}.

To calculate flows in each active region, we first choose an area of
 $15\degr \times 15\degr$ surrounding the active region near disk center and track for 1440 min using the 
surface rotation rate \citep{snod84}. Start and end times of each time series are
provided in  Table~\ref{table1}. 
The spatial sampling of the HMI Dopplergrams is 0.5\arcsec, thus
regions of  $384 \times 384$ pixels were selected.  We use the same tracked data cubes to determine flows in both 
photospheric and sub-photospheric shear layers. The active regions analyzed here are  well-developed and 
moderately large with significant number of sunspots with a spread of about $13\degr$ -- $18\degr$ 
in longitude.   ARs 11139 and 11890 have similar characteristics with low flaring activity while AR 11944
is a much bigger in size, and produced X1.1 and M7.2 class flares, in addition to 5 C-class flares during
the period of analysis. Since gaps in time series introduce uncertainties in flow estimation, the time series 
have been cautiously chosen when the duty cycle  is $\approx$ 100\% and  the ARs are located near disk center
to avoid any influence of systematics in flow determination. Duty cycle and the location of the 
reference image for each timeseries are given in  Table~\ref{table1}.  


\subsection{Photospheric Flows}

Photospheric flows are calculated using the technique of LCT.
This method has been widely used to infer photospheric motion in active regions/sunspots
\citep[][and references therein]{ravindra08}.
In this method, the local velocity in an image is estimated from the displacement of a feature between
two consecutive image $I(x,t)$ and  $I(x,t+\Delta t)$ where $I(x,t)$ represents the reference image as a 
function of position $x$ and time $t$, and the displacement is calculated by cross-correlating pairs of sub-images 
 separated by a uniform time interval $\Delta t$. 

Here, we choose  the object image and reference image that are separated 
 by a 4.5 min interval using a  6.3\arcsec  Gaussian apodizing window function. The corresponding velocities
 in zonal ($x$)- and meridional ($y$)-directions are computed by dividing the displacements by time difference between the two images.  
 Finally, the velocity at each pixel is averaged over 1440 min to determine the long-lived flows in and around 
 active regions.

\subsection{Sub-photospheric Flows}
\begin{figure*}   
   \centerline{
\includegraphics[scale=0.53]{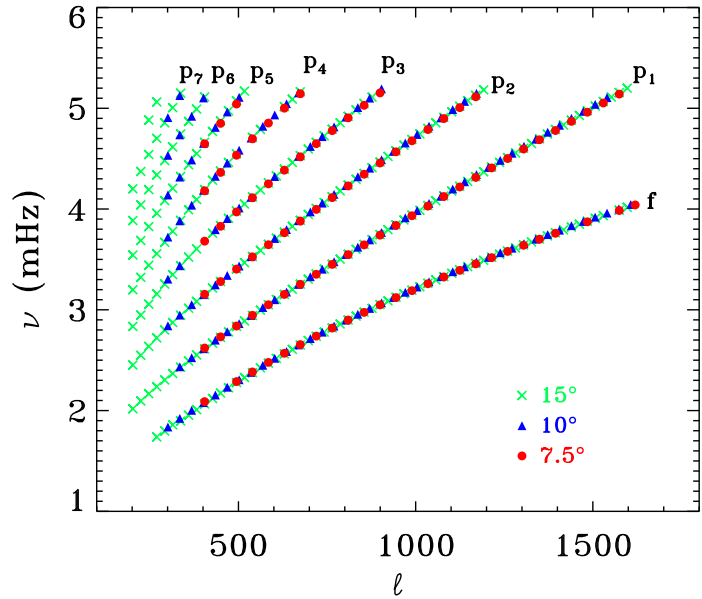}
}
            \caption{Typical $\ell$-$nu$ diagram of the fitted modes at the disk center for different tiles sizes 
            using HMI Dopplergrams. Regions have been tacked for 1440 min on 2013 December 15.
                     }
   \label{modes}
   \end{figure*}

Sub-photospheric flows are calculated using the technique of RD.  In this method, 
high-degree waves propagating in localized areas over the solar 
surface are used to obtain an averaged velocity vector in the  region of interest. It
has been extensively applied to infer sub-surface properties  upto a limited depth. Despite its substantial 
applications, the RD method suffers from the restriction on size of the tiles that restricts the number 
of fitted modes and the depth range covered in the inversion. 
As an example, a typical $\ell$-$\nu$ diagram exhibiting  fitted modes in three different
tile sizes 
using the HMI Dopplergrams is shown in Figure~\ref{modes}. It can be seen that the number of fitted modes decrease with decreasing
tile size. Since a higher value of $\nu/\ell$ denotes the  penetration of the traveling waves at greater depth, the 
smaller tile selection provides reliable results closer to the surface.

 Each tracked area is apodized with a circular function and then a three-dimensional FFT is applied on 
both spatial and temporal direction to obtain a three-dimensional power spectrum. In this
study, the 
corresponding power spectrum is fitted using a Lorentzian profile model \citep{haber00},
\begin{eqnarray}
P (k_x, k_y, \omega) & = & {A \over (\omega - \omega_0 + k_xV_Vx_{Fit}+k_yVy_{Fit})^2+\Gamma^2} \nonumber \\  
&& + {b \over k^3}
\end{eqnarray}
where $P$ is the oscillation power for a wave with a temporal frequency ($\omega$) and
the total wave number $k^2=k_x^2+k_y^2$. There are six parameters to be fitted:
two Doppler shifts ($k_xVx_{Fit}$ and $k_yVy_{Fit}$) for waves propagating in the orthogonal 
zonal and meridional directions, the background power ($b$), the mode
central frequency ($\omega_0$), the mode width ($\Gamma$), and the amplitude ($A$).
Finally, the fitted velocities  ($Vx_{Fit}$ and $Vy_{Fit}$) are inverted using a regularized least square 
(RLS) method to estimate depth dependence of various components of the horizontal velocity ($Vx_{Inv}$ and $Vy_{Inv}$). 

In this paper, we divide the tracked region of $15\degr \times 15\degr$  into a mosaic of overlapping 
tiles where each tile is approximately  $7\degr.5$ $\times$  $7\degr.5$ in size.  
Tiles in the mosaic are spaced by $2\degr.5$ in each 
direction. Thus, there are 49 tiles in the mosaic for each active region. 
Finally, the residual velocity in different regions is calculated by subtracting the velocity in quiet regions at the same 
heliographic location in a nearby Carrington  rotation as discussed in \citet{jain12, jain15}.

\section{Analysis and Results}

\subsection{Flows from RD method}
\begin{figure*}   
   \centerline{
\includegraphics[scale=0.85,angle=90]{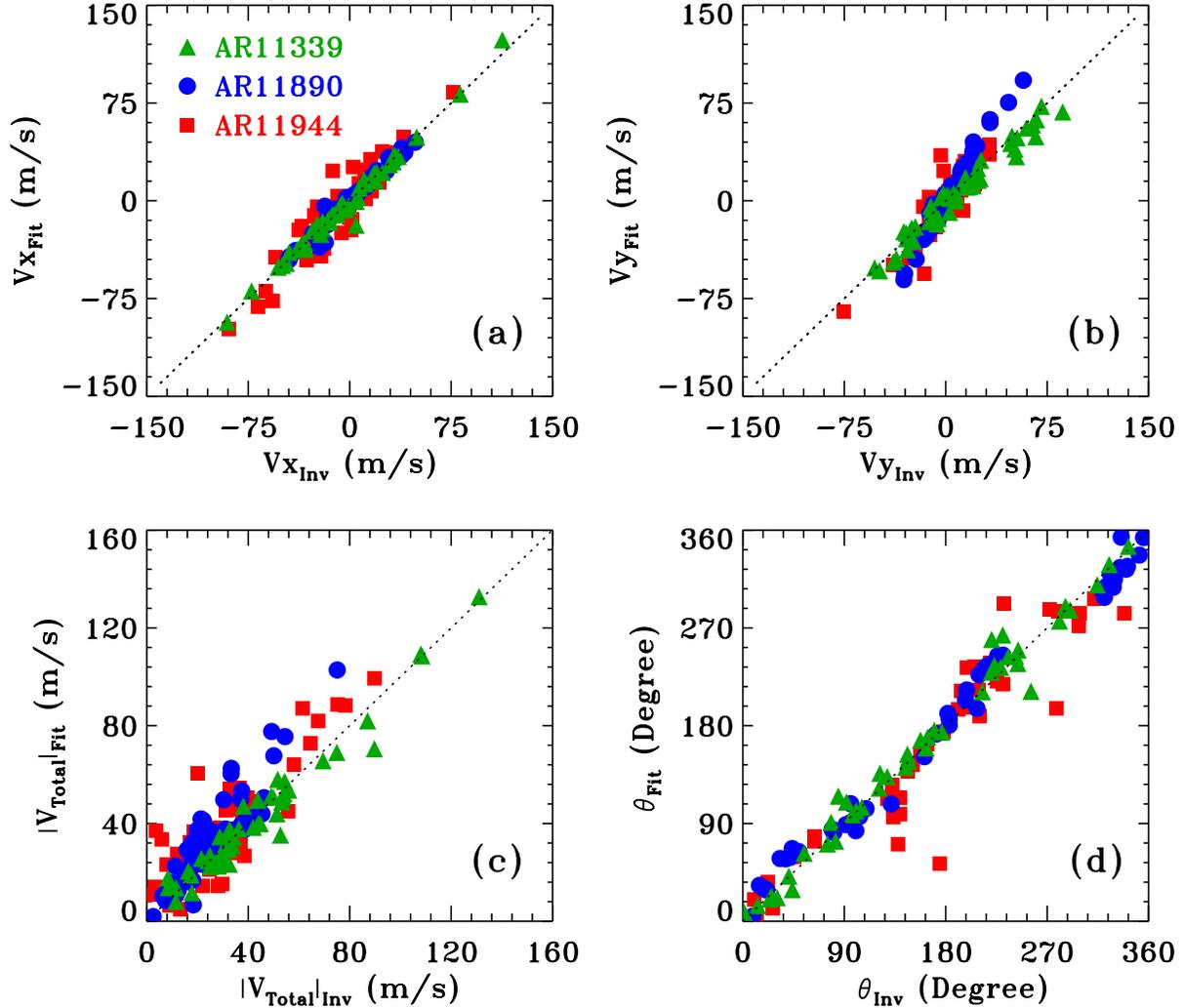}\\
}
            \caption{Comparison of various quantities calculated  using two approaches of ring-diagram analysis;
            (a) zonal-component, (b) meridional-component, (c) total magnitude, and (d) azimuthal angle of the 
            horizontal velocity. Values obtained from the inversion  are plotted on the x-axis and from the 
            fitting are on y-axis. Dotted line in each panel corresponds to
            the ideal scenario where both approaches yield the same results. Statistical uncertainties in velocity estimation
            are smaller than the size of symbols.}
   \label{fit_inv}
   \end{figure*}

We obtain two types of velocities in the RD method; fitted and inverted. In the best scenario, the fitted velocity 
should be comparable to the inverted velocity when proper depth range is taken in to account. In order to infer velocity
in the near-surface shear layer  for comparison with the photospheric values, we compute velocities from both methods: 
(i) calculate velocity residuals of zonal- and meridional-components of the fitted velocity of surface gravity waves   ($f$-modes) only, and
(ii) calculate velocity residuals of inverted velocities in upper 2 Mm.
Note that the surface gravity waves only provides direct measurement
of flow velocity in the layers where $f$ modes have significant amplitude, i.e. the layer spanning about 2 Mm below the surface. 
This approach has been exploited by  several authors in RD analysis, for example,  \citet{ hindman09} studied flows in smaller regions of filaments,
and more recently \citet{rick15} investigated  typical structures associated with magnetic belts and regions of  magnetic activity 
in the outer 1\% of the Sun.  Moreover, the  $f$-mode ridges have also been isolated to construct travel-time maps near the surface in  the 
time-distance technique \citep{ dali07}. In Figure~\ref{fit_inv}, we demonstrate the direct 
comparison between fitted and inverted velocities.  Here, we show scatter plots of four quantities, i.e. $V_x$, $V_y$, total magnitude 
($|V_{Total}| = (V_x^2 + V_y^2)^{1/2}$),  and  the azimuthal angle ($\theta$) of the horizontal  velocity. 
In all cases, we show the ideal scenarios by the dotted lines where fitted and inverted values should be in agreement. 

We statistically test the correspondence between both methods by calculating Pearson  
($r_P$) and Spearman   ($r_S$) correlation coefficients. While $r_P$ is used to measure the linear association
between two variables, $r_S$  measures the extent of association that may exist between two 
series of ranks for the same set of variables. The linear correlation is calculated 
using the following expression;

\begin{equation}
r_P   =  { \sum_{i}{{u}_i{v}_i} \over ({\sum_{i}{|{u}_i|^2} {\sum_{i}|{v}_i|^2}})^{1/2} },
\end{equation}
where ${u}_i$ and ${v}_i$ are two scalar variables.   When there are no tied ranks in either column of data, 
the $r_S$ can be simplified to 

\begin{equation}
r_S   =  1- {{ 6\sum_{i=1}^n{d_i^2}} \over {n^3-n}}
\end{equation}
where  $d_i$ the difference between ranks for observation $i$, and $n$ is the sample size.

We find a good agreement between fitted and inverted values. The calculated  values of $r_P$ 
for $V_x$, $V_y$, $|V_{Total}|$ and  $\theta$ are 0.99, 0.98, 0.91 and 0.87, respectively  
while corresponding $r_S$ are 0.98, 0.99, 0.84, 0.97. It is also noticed that the individual 
components from two approaches are better correlated than the total magnitude and
azimuthal angle. We further notice that the fitted zonal and meridional components for AR 11339 
are much closer to the inverted values as compared to other two ARs. The maximum deviation is
seen in the meridional component of AR 11890 where fitted values are higher than those obtained
from inversion. 

\subsection{Flows from LCT Method}
\begin{figure*}   
   \centerline{
\includegraphics[scale=.9]{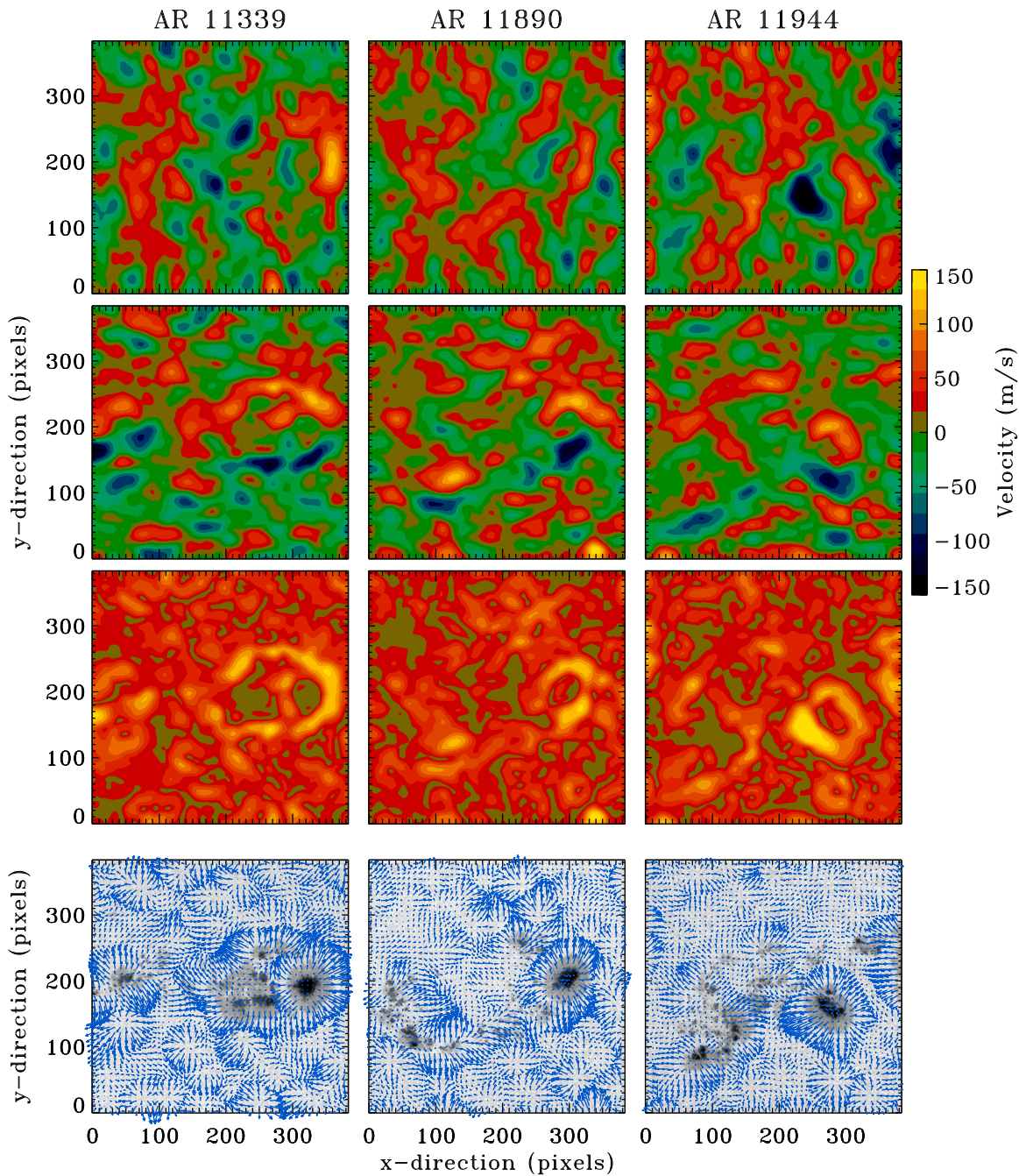}
}

            \caption{Maps of photospheric (top row) zonal flow, (2nd row) meridional flow, (3rd row)
            total horizontal flow, and (bottom row) flow vectors over-plotted on the HMI continuum image,
           from the LCT method for three 
            active regions: AR 11339 (left), AR 11890 (middle), and AR 11944 (right). The positive/negative
            values are for west/east zonal flows and north/south meridional flows.
            Background images in bottom row 
            are same as in Figure~\ref{regions}. }
   \label{lct}
   \end{figure*}

We show, in  Figure~\ref{lct}, the maps of zonal and  meridional components of horizontal 
velocity at the surface for each AR using the LCT method. These maps display values between +30 to $-$30 m/s
in general except for some areas of higher values. A closer examination of these maps with Figure~\ref{regions}
clearly indicates that the higher values are obtained at the locations of sunspots. We also include the 
maps of magnitude (3rd row from the top) and the direction (bottom row) of total horizontal velocity in
the same figure. In these maps, we find an excess velocity in  penumbrae of large sunspots. 
We also obtain an outflow in penumbrae  and the 
inflow in umbrae of all major sunspots in active regions. These findings are in agreement with earlier results.  

\subsection{Comparison between Flows from RD and LCT Methods: Zonal and Meridional Components}

\begin{figure}   
   \centerline{
\includegraphics[scale=0.78]{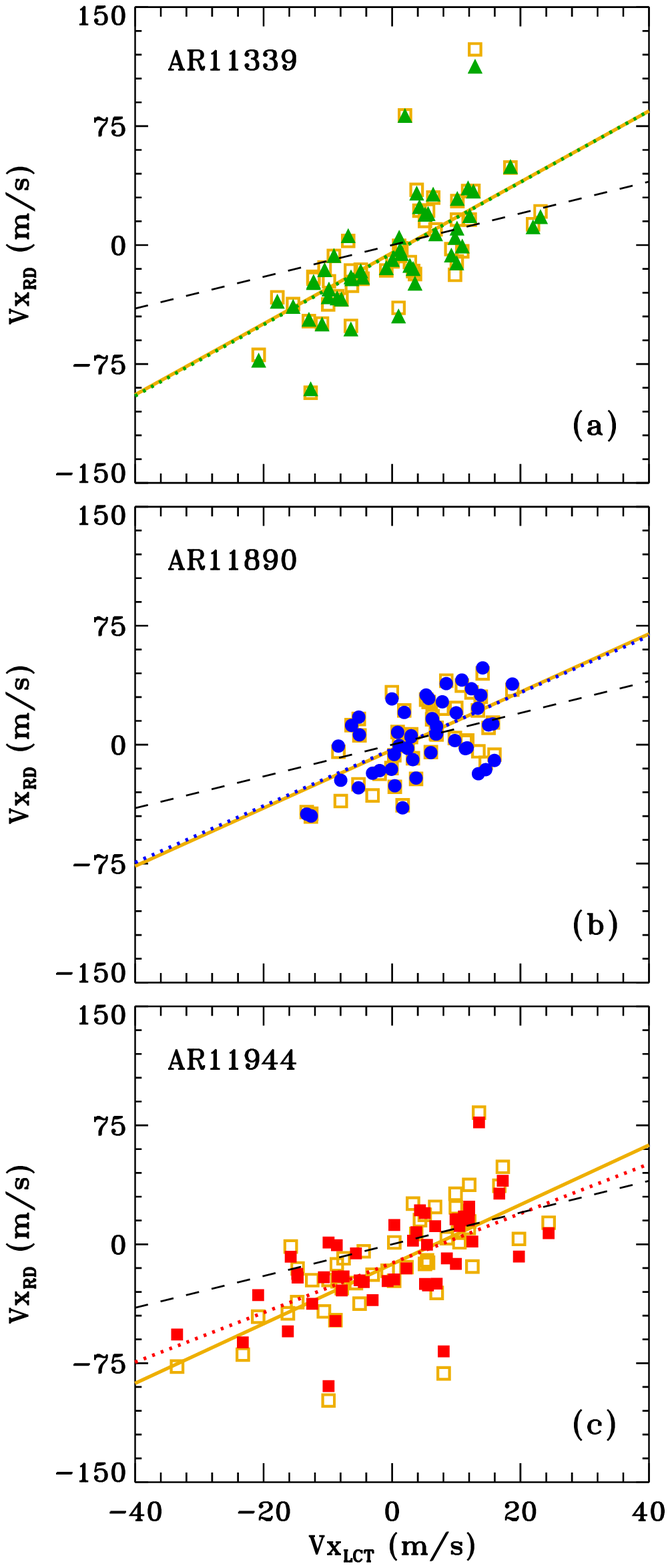}\\
}
            \caption{Comparison between zonal-components of  photospheric flows from LCT ($Vx_{LCT}$)and sub-photospheric flows
            near surface from ring-diagram technique ($Vx_{RD}$) in three active regions; (a) AR 11339, (b) AR11890, 
            and (c) AR11944. 
            Sub-photospheric flows calculated using velocities of  fitted f modes are shown by open yellow squares  while inverted
            flows are shown by the filled symbols,  and their linear fits  are shown by solid and dotted lines, respectively.
Dashed lines represent the ideal scenario for both velocities, $Vx_{LCT}$ and $Vx_{RD}$. Uncertainties in velocity estimation
            are smaller than the size of symbols.
  }
   \label{zonal}
   \end{figure}

\begin{figure}   
   \centerline{
\includegraphics[scale=0.78]{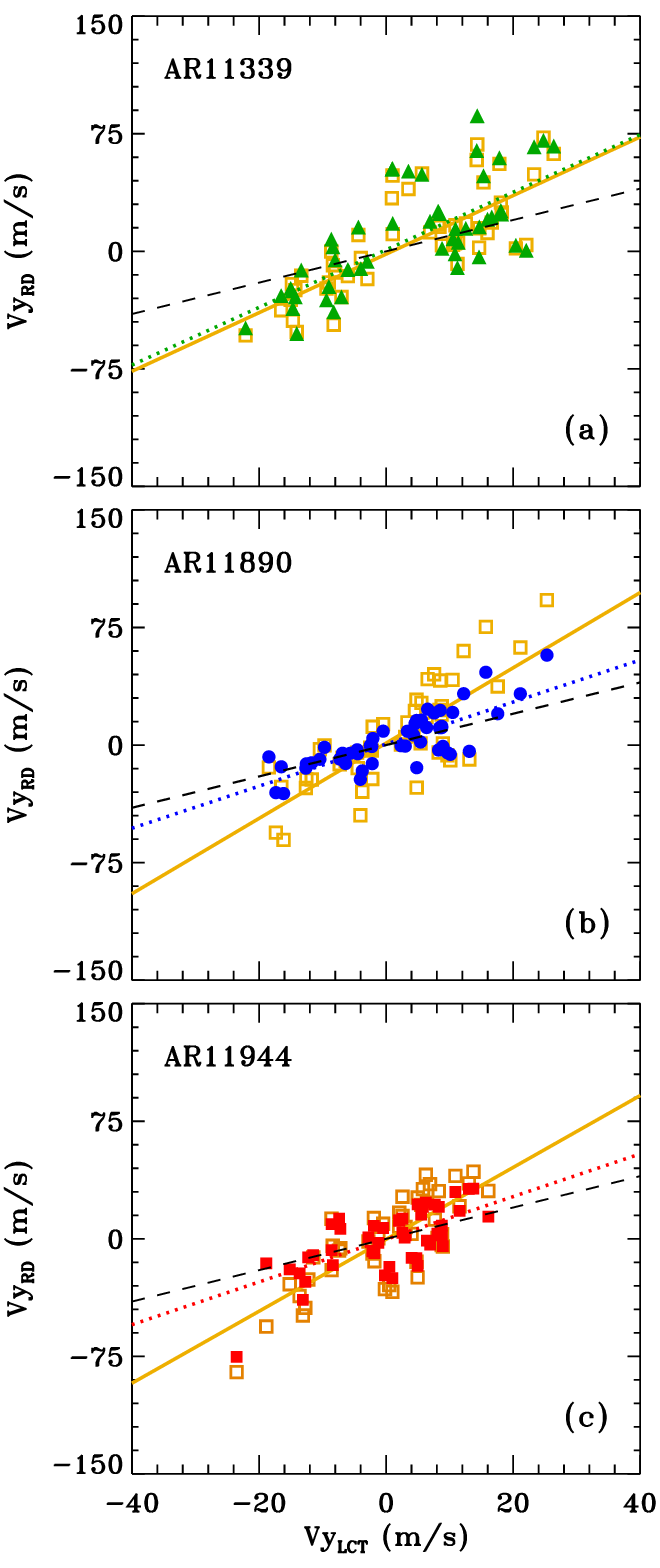}\\
}
            \caption{Comparison between meridional-components of  photospheric flows from LCT ($Vy_{LCT}$) and sub-photospheric flows
            near the surface from ring-diagram technique ($Vy_{RD}$) in three active regions; (a) AR 11339, (b) AR11890, 
            and (c) AR11944. 
            Sub-photospheric flows calculated using velocities of  fitted f modes are shown by open yellow squares  while inverted
            flows are shown by the filled symbols,  and their linear fits  are shown by solid and dotted lines, respectively.
Dashed lines represent the ideal scenario for both velocities, $Vy_{LCT}$ and $Vy_{RD}$. Uncertainties in velocity estimation
            are smaller than the size of symbols.}
   \label{meridional}
   \end{figure}

As discussed earlier, the flows from LCT and RD methods are generally calculated at two different spatial scales. While LCT is able to
track changes in velocity fields of small elements on the surface, the RD method provides estimates of average velocity only
for regions that are much bigger in size. In order to examine the similarities/dissimilarities between them, 
the velocity components from LCT have been averaged over the same grid as in the RD method. This allows us to perform one-to-one 
comparison between flows from both methods. Figures~\ref{zonal} and \ref{meridional} exhibit scatter plots of $x$- and $y$- components
from the LCT method vs. those obtained from both approaches of the RD method. We notice a visible difference in their magnitudes, 
the values obtained from the RD method are larger than those from the LCT method. The solid and dotted lines in 
these figures correspond to the best linear-fit between LCT and two values from the RD  while the dashed lines are drawn 
to depict the same values from all methods. The linear fits are obtained from the least square method; the errors in LCT  
are considered to be zero while in RD these are the statistical uncertainties in velocity determination.

In order to quantify the similarities and/or differences in  computed zonal and meridional components from different methods, 
we provide results of the statistical analysis in Table~\ref{table2}. In most cases, the 
correlation coefficients are higher than 0.60. We note that the LCT values are better correlated with the RD fitted values than those from the 
inversion. Also, the meridional component has higher correlation than the zonal component. This is in contrast to  the results from
time-distance analysis where  the zonal component of the flow near surface is found to be better 
correlated with the surface measurements \citep{liu13,svanda07}. 
We also compute slopes of the best-fitted lines for individual active regions and by combining all data points together. The slopes differs
significantly from the ideal value 1.0. However, we also obtain similar slopes for both fitted and inverted values for AR 11339. This is 
similar to the results shown in Figure~\ref{fit_inv} where both approaches from RD for AR 11339 yield the similar values.  We further 
interpret these slopes as the scaling factors  factors between the RD and LCT values. In general, fitted values have higher slopes than the 
inverted values indicating larger deviation from the LCT values, and also higher correlation coefficients. Quantitatively, both zonal 
and meridional components
in LCT are smaller by a factor of 2 as compared to the fitted velocities and by a factor of 1.8 and 1.5 against inverted zonal
and meridional velocities, respectively. 

\subsection{Comparison between Total Horizontal Flows from RD and LCT Methods}

\begin{figure}   
   \centerline{
\includegraphics[scale=.55]{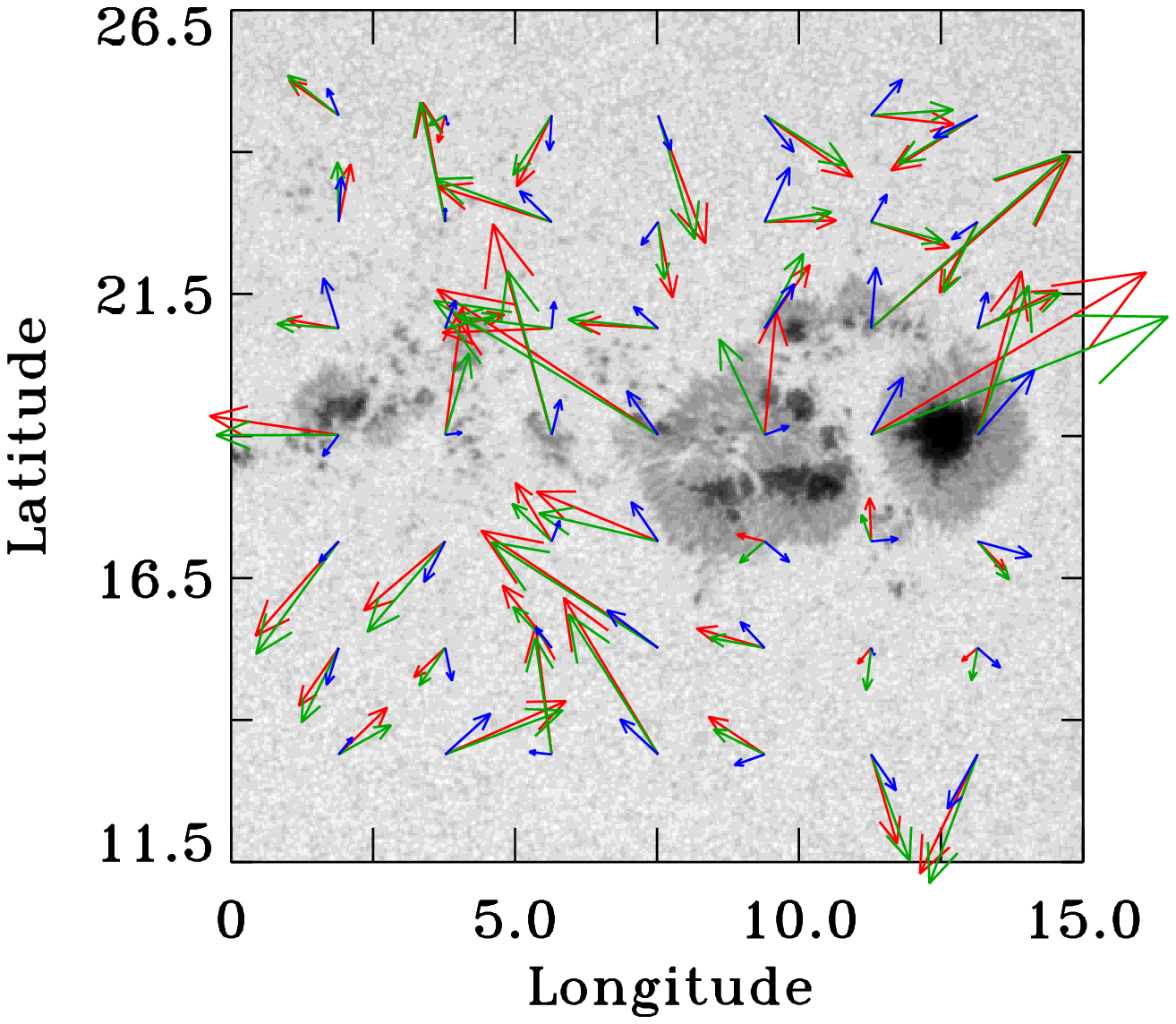}
}
  \centerline{
\includegraphics[scale=.55]{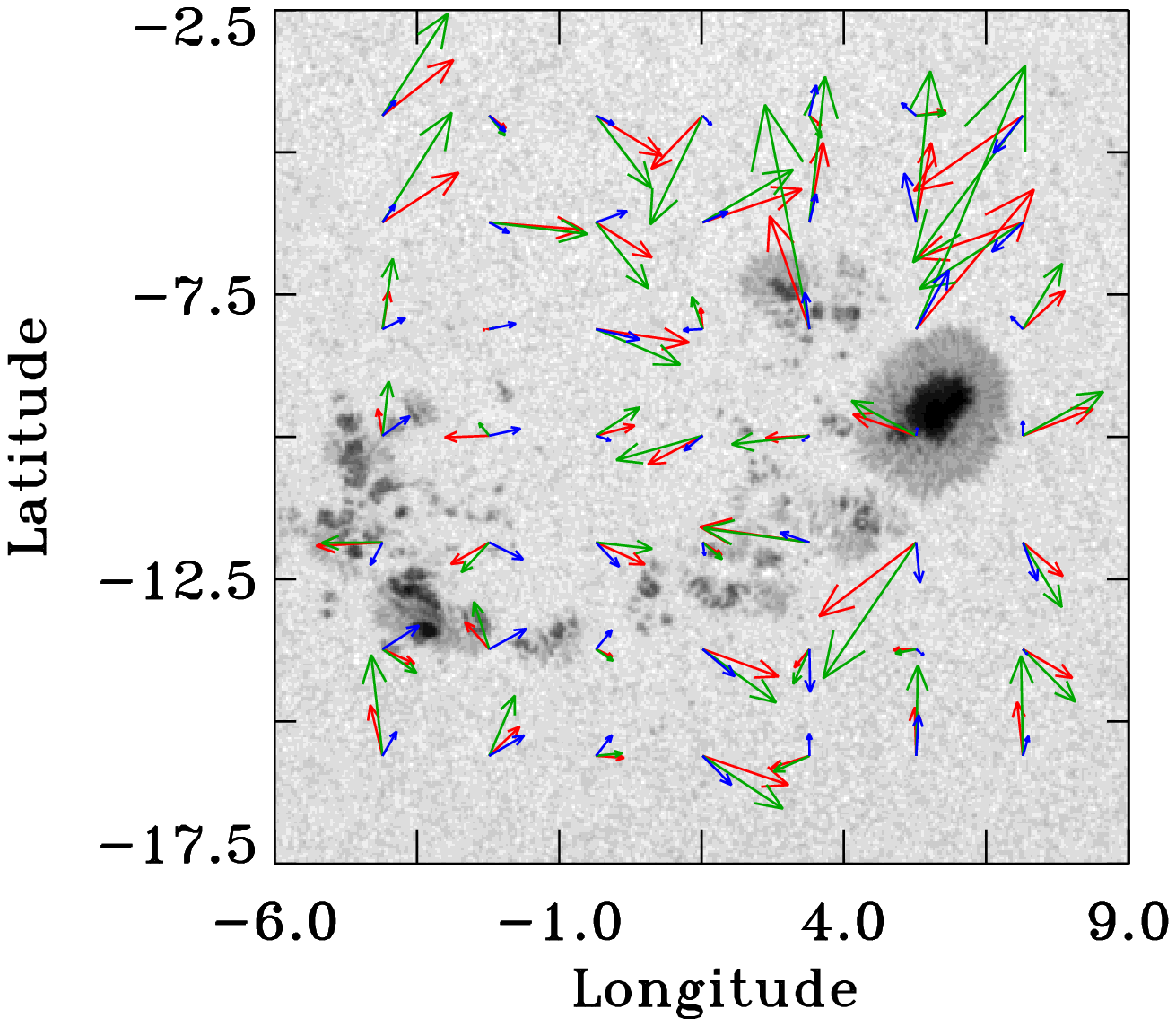}
}
  \centerline{
\includegraphics[scale=.55]{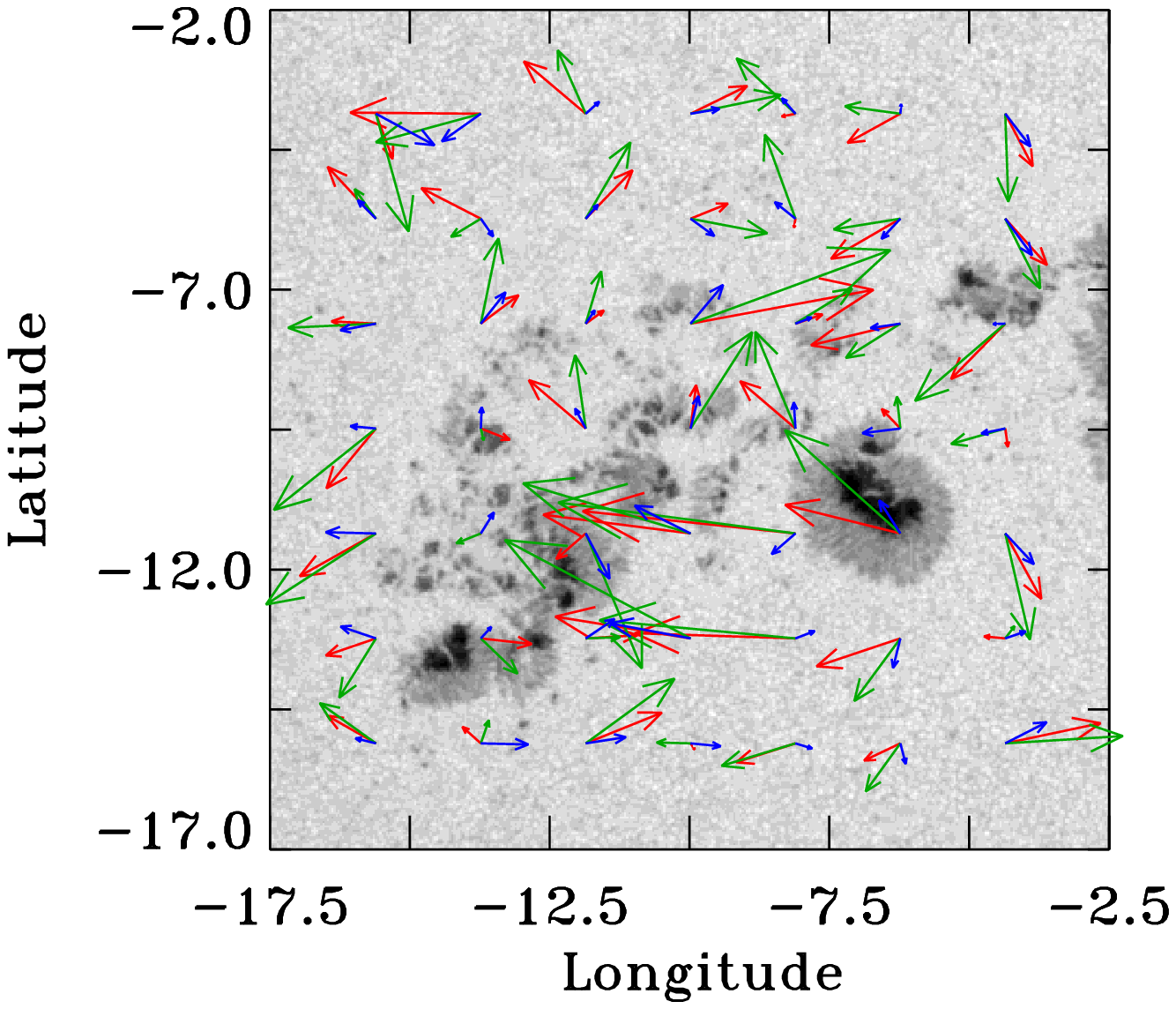}
              }
            \caption{Comparison of  horizontal flows -- photospheric from LCT (blue), sub-photospheric
            from f-modes (green) and sub-photospheric inverted velocities (red) in AR 11339 (top), 
            AR 11890 (middle), and AR 11944 (bottom). 
}
   \label{total}
   \end{figure}

We display in Figure~\ref{total} the total horizontal velocity for all three active regions. In addition to the magnitudes, here we also 
compare the directions of velocity vectors. In order to understand the variation in velocity vectors from region-to-region, 
we over-plot these vectors on the HMI continuum images. We find that the large values are obtained for tiles in the vicinity 
of big sunspots that host the reservoir of strong magnetic fields. A visual inspection of these plots 
hints for the agreement between different methods, although the magnitudes from LCT are significantly smaller than the 
RD values.  Major discrepancies in the direction are found for tiles with large sunspots, however there are only a very
few tiles where flows from both methods point to the different directions. A detailed quantative study is needed to investigate the 
correlation between the differences in direction using a large database of active regions and also by including
the inclination angle determined from the vector magnetic field data. 

We further quantity the degree of agreement in these vectors by analyzing their local characteristics, i.e. the vector 
magnitude and direction at each grid point. We use all three total horizontal velocity vectors, i.e.,  
$V_{Fit}$, $V_{Inv}$, and $V_{LCT}$ and  compute the vector correlation coefficient 
($C_{vec}$) and the Cauchy-Schwarz inequality  ($C_{CS}$) as described below;

\begin{equation}
C_{vec}   =  { \sum_{i}{\mathbf{U}_i\cdot\mathbf{V}_i} \over ({\sum_{i}{|\mathbf{U}_i|^2} {\sum_{i}|\mathbf{V}_i|^2}})^{1/2} },
\end{equation}

\begin{equation}
C_{CS}   =   {{ {1} \over {M}}  \sum_{i} {{\mathbf{U}_i\cdot{\mathbf{V}_i} \over {|\mathbf{U}_i| |\mathbf{V}_i|}}  }  
          =  {{1} \over {M}}{\sum_{i} cos\theta_i   } }   \\
\end{equation}
where $\mathbf{U}_i$ and $\mathbf{V}_i$ are two velocity vectors, $\theta_i$ the angle between them, and M the total number of vectors. The $C_{vec}$ 
is equivalent to the correlation coefficients for scalar functions but for the vector quantities while the $C_{CS}$ a measure 
of the angular difference between two vector fields; it is 1 when the fields are parallel and $-$1  for anti-parallel fields. Computed   $C_{vec}$ and $C_{CS}$
are listed in  Table~\ref{table3} where $r_P$  coefficients are also included for comparison. We find higher values for $C_{vec}$ as compared 
the  $r_P$. Obtained positive higher values of $C_{CS}$ also indicate that these velocity
vectors  point to the same direction. We again find better agreement between LCT and fitted velocities over the LCT and inverted velocities.
This might be due to the way these quantities are computed as the  inversion is performed on the fitted velocities that may add another set of 
uncertainties to the inverted velocities.
Although due to the limitations on spatial scale in the RD method, we can not
compute flows confined to penumbral and umbral regions separately, a close correspondence between LCT and RD methods
clearly implies similar flows are inferred from both the methods in these areas.

\section{Summary}
 We have measured zonal and meridional components of the horizontal velocity near the surface in three active regions.
 Velocities  in upper 2 Mm of the sub-photospheric layer are calculated using the ring-diagram technique of helioseismology while the  
 local correlation tracking of surface features is applied to estimate flows in the photosphere. Although both methods employ different 
 spatial scales, we find positive and significant correlation  between the individual components 
 of the flows. This  clearly indicates that despite the absorption of acoustic power in active regions, the overall trends in 
 flows calculated from the helioseismology are comparable to the surface measurements. However, the magnitudes of velocity in 
 both methods are  significantly different. The velocity from LCT method is  smaller by a factor of 2 (as a consequence of smoothing)
 and the median difference between direction of corresponding vectors is $49\degr$.
 Further, the magnitudes of 
 fitted and inverted velocities from
 the ring-diagram method are comparable implying that the inversion technique used in the ring-diagram analysis (RLS in this case) provides
 reliable estimates of the inferred flows. 
 
 
\acknowledgments

{\it SDO} data courtesy of SDO (NASA) and the HMI and AIA consortium. This work was partially supported 
by NASA grant NNH12AT11I and NSF Award 1062054 to the National Solar Observatory. The ring-diagram 
analysis was carried out using the NSO/GONG ring-diagram pipeline. This work was performed 
under the auspices of the SPACEINN Framework of the European Union (EU FP7).  

\newpage
\bibliography{Jain}

\begin{deluxetable}
{ccccccccccc}
\tabletypesize{\scriptsize} 
\tablecaption{Details of Data Used in this Study. \label{table1}  }
\tablewidth{0pt}
\tablehead{
 \multicolumn{3}{c} {Active Region}    &&   \multicolumn{2}{c} {Timeseries} & Duty &&   \multicolumn{2}{c} {Location\tablenotemark{c}} \\
\cline{1-3} \cline{5-6} \cline{9-10}
 {Number} & Type\tablenotemark{a}  &Area\tablenotemark{b} & &     {Start}   & {End} &{Cycle} & &  {Longitude}   & {Latitude}    
}
\startdata
NOAA 11339&$\beta\gamma$ &940&& 2011 Nov 08 & 2011 Nov 09 & 99.84\%&& 7\degr.5 & 19\degr.0  \\
&&&&12:00:00 UT & 11:59:15 UT &&&\\
\\
NOAA 11890&$\beta\gamma\delta$&920 && 2013 Nov 08 & 2013 Nov 09 &99.95\% && 1\degr.5 & -10\degr.0 \\
&&&&10:00:00 UT & 09:59:15 UT &&&\\
\\
NOAA 11944&$\beta\gamma\delta$&1540 && 2014 Jan 06 & 2014Jan 07 & 99.63\% &&  -10\degr.0 & -9\degr.5 \\
&&&&23:30:00 UT & 23:29:15 UT &&&
\enddata
\tablecomments{}
\tablenotetext{a}{Magnetic configuration of the spots in the Mt. Wilson system}
\tablenotetext{b}{Area of spots in millionths of the visible hemisphere}
\tablenotetext{c}{The location are listed for the central pixel of reference image in $15\degr \times 15\degr$ tile.}

\end{deluxetable}

\begin{deluxetable}
{ccccccccccccccccc}
\tabletypesize{\scriptsize} 
\tablecaption{Statistical Analysis between Flows from Local Correlation
Tracking (LCT) and Ring-diagram (RD) Methods.\label{table2}  }
\tablewidth{0pt}
\tablehead{
AR  & RD  &   \multicolumn{4}{c} {Zonal} &&   \multicolumn{4}{c} {Meridional}  \\
\cline{3-6} \cline{8-11} 
Number & Flow\tablenotemark{a}  & {$r_P$} & {$r_S$} &{$P_S$} & {Slope\tablenotemark{b} } & & {$r_P$} & {$r_S$} &{$P_S$}& {Slope\tablenotemark{b} }
}
\startdata
11339 & Fit  &  0.69 & 0.79 &10$^{-11} $ & 2.24$\pm$0.03  & &  0.78 & 0.77 & 10 $^{-11} $ & 1.87$\pm$0.02  \\
     &  Inv  &  0.71 & 0.82 &10$^{-13} $ & 2.25$\pm$0.04  & &  0.73 & 0.73 & 10 $^{-09} $ & 1.84$\pm$0.03\\
\\
11890 & Fit  &  0.62 & 0.54 &10$^{-06} $ & 1.83$\pm$0.04  & &  0.78 & 0.74 &10 $^{-09} $ & 2.51$\pm$0.03\\
      & Inv  &  0.58 & 0.51 &10$^{-05} $ & 1.79$\pm$0.04  & &  0.79 & 0.75 &10 $^{-10} $ & 1.34$\pm$0.03 \\
\\
11944 & Fit  &  0.69 & 0.74 & 10 $^{-09} $ & 1.88$\pm$0.02  & &  0.79 & 0.73 & 10 $^{-09} $ & 2.29$\pm$0.04\\
     &  Inv  &  0.65 & 0.67 & 10$^{-07} $ & 1.56$\pm$0.03  & &  0.71 & 0.66 & 10 $^{-07} $ & 1.36$\pm$0.03 \\
\\
All  & Fit  &  0.68 & 0.73 &10$^{-25} $ & 2.01$\pm$0.02  & &  0.78 & 0.76 & 10 $^{-28} $ & 2.11$\pm$0.02 \\
     &  Inv  &  0.66 & 0.70 &10$^{-22}$ & 1.87$\pm$0.02  & &  0.74 & 0.74 & 10 $^{-26} $ & 1.51$\pm$0.02 \\
\enddata
\tablecomments{}

\tablenotetext{a}{Type of RD method to calculate flows.}
\tablenotetext{b}{Slopes are calculated by performing the least square fits between LCT and RD velocities;
the errors in LCT  are considered to be zero while in RD these are the statistical uncertainties.}
\end{deluxetable}


\begin{deluxetable}
{cccccc}
\tabletypesize{\scriptsize} 
\tablecaption{Merit Analysis of Total Horizontal Velocity. \label{table3}  }
\tablewidth{0pt}
\tablehead{
Velocity Vectors &Metrics of Merit&  AR 11339  & AR11890 &AR 11944 &All\\ 

}
\startdata
$V_{LCT}$ \& $V_{Fit}$&$r_P$ & 0.55& 0.43 & 0.54& 0.53\\
 & $C_{vec}$ & 0.73& 0.72& 0.71 & 0.72\\
 & $C_{CS}$  & 0.72&  0.67& 0.65& 0.67\\
 \\
 $V_{LCT}$ \& $V_{Inv}$&$r_P$ &0.50 & 0.43&  0.46& 0.47\\
 & $C_{vec}$ & 0.72&  0.66&  0.64 & 0.69\\
 & $C_{CS}$  & 0.71&  0.61& 0.60&0.62\\

\enddata

\end{deluxetable}

\end{document}